\documentclass{article}
\usepackage{amssymb}

\usepackage{graphicx}
\usepackage{amsmath}

\newenvironment{proof}[1][Proof]{\textbf{#1.} }{\ \rule{0.5em}{0.5em}}
\input{tcilatex}

\begin{document}

\title{Holevo-ordering and the continuous-time limit for open Floquet dynamics}
\author{John Gough \\
%EndAName
Department of Computing \& Mathematics\\
Nottingham-Trent University, Burton Street,\\
Nottingham NG1\ 4BU, United Kingdom.\\
john.gough@ntu.ac.uk}
\date{}
\maketitle

\begin{abstract}
We consider an atomic beam reservoir as a source of quantum noise. The atoms
are modelled as two-state systems and interact one-at-a-time with the
system. The Floquet operators are described in terms of the Fermionic
creation, annihilation and number operators associated with the two-state
atom. In the limit where the time between interactions goes to zero and the
interaction is suitably scaled, we show that we may obtain a causal (that
is, adapted) quantum stochastic differential equation of
Hudson-Parthasarathy type, driven by creation, annihilation and conservation
processes. The effect of the Floquet operators in the continuous limit is
exactly captured by the Holevo ordered form for the stochastic evolution.
\end{abstract}

\section{Introduction}

Periodically kicked quantum systems are described dynamically by applying a
unitary $V$, called the Floquet operator, every $\tau $ seconds. In an open
systems model, $V$ will be a unitary on a Hilbert space $\frak{h}_{S}\otimes 
\frak{h}_{R}$ where $\frak{h}_{S}$ is the system's state space and $\frak{h}%
_{R}$ is the state space for the environment at that particular time.
Averaging over the environment leads to a dissipative reduced dynamics on $%
\frak{h}_{S}$. Here we address the question of an system being periodically
kicked by independent environments and consider the continuous time limit $%
\tau \rightarrow 0$. To obtain an limit open dynamics we must re-scale the
Floquet operators appropriately.

The approach which best captures the limit dynamics is that of the
``time-ordered exponentials'' introduced by Holevo \cite{Holevo}. This
theory essentially deals with quantum stochastic Floquet operators and is
equivalent to the usual Hudson-Parthasarathy approach \cite{HP}. We prefer
the terminology Holevo ordered from for the former and Wick ordered form for
the latter. This is because time ordered exponentials appear as Dyson series
expansions: here the chronologically ordered terms lead to a
Weyl-Stratonovich theory as opposed to the Wick-It\^{o} of Hudson and
Parthasarathy \cite{GREP}. Previously, we have established a quantum central
limit for time-ordered exponentials involving reservoir fields $%
a_{t}^{+}\left( \lambda \right) ,a_{t}^{-}\left( \lambda \right) $
satisfying commutation relations $\left[ a_{s}^{-}\left( \lambda \right)
,a_{t}^{+}\left( \lambda \right) \right] =\frac{1}{\lambda ^{2}}G\left( 
\frac{t-s}{\lambda ^{2}}\right) $, where $\lambda $ was a small parameter,
and emission, absorption and scattering where present \cite{G1JMP}. This is
interpreted as Markovian limit where the auto-correlation time $\tau
\varpropto \lambda ^{2}$ vanishes. The fields $a_{t}^{\pm }\left( \lambda
\right) $ are actually superpositions of creation/annihilation fields for
fixed momenta states modulated by a $t$-dependent phase component: the
Markovian approximation is then an infinite bandwidth limit.

In the present situation, the fields are in discrete time and are
Kronecker-delta correlated. In this sense they are already discrete white
noises. However we use the same strategy of anticipating the limit in terms
of suitably scaled collective operators leading to a quantum central limit:
see \cite{CH},\cite{GvW} and chapter II of \cite{MeyerQP}. We have the
advantage here that the dynamical updates involve operators independent of
past state of the reservoir and so we avoid the finite-memory features
appearing in Markovian approximations.

The limit we consider has been used to describe the open dynamics of a laser
mode interacting with an atomic beam reservoir \cite{Kistetal}. This problem
has also been studied recently by Attal and Pautrat \cite{AttalPautrat} and
we obtain similar results to theirs. They, however, investigate the vacuum
limit and use the Guichardet's representation of Fock space processes and
the toy-Fock approximation to Fock space \cite{MeyerQP}. For our purposes,
we find the connection to Holevo's formalism the most transparent. We
conclude with a construction of non-vacuum limits.

\section{Model For an Atomic Beam Reservoir}

\subsection{Open Floquet Dynamics}

Let $V$ be a unitary operator on the Hilbert space $\frak{h}_{S}\otimes 
\frak{h}_{R}$\ where $\frak{h}_{S}$ is states space for a system of interest
and $\frak{h}_{R}$ the state space for its current environment. If we fix a
reference density operator $\varrho _{R}$ for the environment, then a
completely positive map, $\Xi $, on the algebra $B\left( \frak{h}_{S}\right) 
$ of bounded system operators is determined by 
\begin{equation}
\text{tr}_{\frak{h}_{S}}\left\{ \varrho _{S}\,\Xi \left( X\right) \right\} =%
\text{tr}_{\frak{h}_{S}\otimes \frak{H}_{R}}\left\{ \varrho _{S}\otimes
\varrho _{R}\,V^{\dagger }\left( X\otimes 1\right) V\right\}
\end{equation}
If the environment was ignored, then $V$ would be referred to as a Floquet
operator, particularly, when applied repeatedly. In such cases $\Xi \left(
X\right) \equiv V^{\dagger }XV$ is a closed, and so non-dissipative,
evolution.

Our aim is to study \textit{open} Floquet systems and for simplicity we
shall assume that the repeated applications of the Floquet operator involve
copies of the same unitary however with different, independent environments.

We consider a repeated interaction strategy as a discrete-time open
dynamics. At times $t=\tau ,2\tau ,3\tau ,\dots $ we have an application of
a copy of the Floquet operator $V$. Let $\frak{h}_{R,k}$ be the state space
at time $t=k\tau $ - this will be a copy of $\frak{h}_{R}$\ - then we are
interested in the Hilbert spaces 
\begin{equation}
\frak{H}_{R}^{t]}=\bigotimes_{k=1}^{\left\lfloor t/\tau \right\rfloor }\frak{%
h}_{R,k},\qquad \frak{H}_{R}^{(t}=\bigotimes_{k=\left\lfloor t/\tau
\right\rfloor +\tau }^{\infty }\frak{h}_{R,k},\qquad \frak{H}_{R}^{\tau }=%
\frak{H}_{R}^{t]}\otimes \frak{H}_{R}^{(t}
\end{equation}
where $\left\lfloor x\right\rfloor $ means the integer part of $x$. (We fix
a vector $e_{0}\in \frak{h}_{R}$ and use this to stabilize the infinite
direct product.) We refer to $\frak{H}_{R}^{t]}$ and $\frak{H}_{R}^{(t}$\ as
the past and future reservoir spaces respectively.

The Floquet operator to be applied at time $t=k\tau $ will be denoted $V_{k}$
and acts on the joint space $\frak{h}_{S}\otimes \frak{H}_{R}^{\tau }$ but
has non-trivial action only on the factors $\frak{h}_{S}$ and $\frak{h}%
_{R,k} $. The unitary operator $U_{t}^{\left( \tau \right) }$ describing the
evolution from initial time to time $t$ is therefore 
\begin{equation}
U_{t}^{\left( \tau \right) }=V_{\left\lfloor t/\tau \right\rfloor }\cdots
V_{2}V_{1}
\end{equation}
It acts on $\frak{h}_{S}\otimes \frak{H}_{R}^{\tau }$ but, of course, has
trivial action on the future reservoir space. The same is true of the
discrete time dynamical evolution of observables $X\in B\left( \frak{h}%
_{S}\right) $ given by 
\begin{equation}
J_{t}^{\left( \tau \right) }\left( X\right) =U_{t}^{\left( \tau \right)
\dagger }\left( X\otimes 1\right) U_{t}^{\left( \tau \right) }\text{.}
\end{equation}

We therefore have the difference equation 
\begin{equation}
\frac{1}{\tau }\left( U_{\left\lfloor t/\tau \right\rfloor +\tau }^{\left(
\tau \right) }-U_{\left\lfloor t/\tau \right\rfloor }^{\left( \tau \right)
}\right) =\left( V_{\left\lfloor t/\tau \right\rfloor +\tau }-1\right)
U_{\left\lfloor t/\tau \right\rfloor }^{\left( \tau \right) }.
\end{equation}
Our objective is to obtain a (quantum) stochastic differential equation for
limiting situation $\tau \rightarrow 0$. To this end, we require a reference
state for the reservoir and we choose the\ pur state determined by the
vector $\Phi ^{\tau }$ on $\frak{H}_{R}^{\tau }$ given by 
\begin{equation*}
\Phi ^{\tau }=e_{0}\otimes e_{0}\otimes e_{0}\otimes e_{0}\cdots
\end{equation*}
and, since $e_{0}$ will typically be identified as the ground state on $%
\frak{h}_{R}$, we shall call $\Phi ^{\tau }$ the vacuum vector for the
reservoir.

The situation can be describe alternatively as follows. The Hamiltonian
describing the combined system and reservoir is the formal operator on $%
\frak{h}_{S}\otimes \frak{H}_{R}^{\tau }$ given by 
\begin{equation}
H_{t}^{\left( \tau \right) }=\sum_{k=1}^{\infty }\delta \left( t-k\tau
\right) \mathcal{H}_{k}^{\left( \tau \right) }
\end{equation}
where $\mathcal{H}_{k}^{\left( \tau \right) }$ acts non-trivially only on
the factors $\frak{h}_{S}$ and $\frak{h}_{R,k}$. The Floquet operators are
then 
\begin{equation}
V_{k}=\exp \left\{ -i\tau \mathcal{H}_{k}^{\left( \tau \right) }\right\} .
\end{equation}
Note that we include a dependence on the time-scale parameter $\tau $ in $%
\mathcal{H}_{k}^{\left( \tau \right) }$ as we shall require some control
over the interaction in the limit $\tau \rightarrow 0$.

\subsection{Atomic Beam Reservoirs}

One situation that we can model in this way is when the reservoir $R$ is a
beam consisting of discrete atoms, each having state space $\frak{h}_{R}$.
The system might be a photon mode inside a cavity. The atoms pass through
the cavity in a regular sequence and its is assumed that the atom-mode
interaction takes place over a time period $\tau _{\mathrm{int}}$ shorter
than the time $\tau $ taken for a single atom to pass through the cavity:
therefore the atoms are independent and at any time at most one atom
interacts with the mode. We shall therefore assume that the interaction is
instantaneous - that is, the system receives a ``kick'' from each atom.

For simplicity, the atoms are taken to be just two-level atoms with ground
state $e_{0}$ and excited state $e_{1}$. The transition operator from the
ground state to the excited state of the $k-$th atom is $\sigma _{k}^{+}$
and is the ampliation of $\sigma ^{+}=\left| e_{1}\right\rangle \left\langle
e_{0}\right| $: its adjoint is denoted as $\sigma ^{-}=\left|
e_{0}\right\rangle \left\langle e_{1}\right| $ and the ampliation by $\sigma
_{k}^{-}$. The operators $\sigma _{k}^{\pm }$ are Fermionic variables and
satisfy the anti-commutation relation 
\begin{equation}
\{\sigma _{k}^{\pm },\sigma _{k}^{\pm }\}=0,\text{\ \ }\{\sigma
_{k}^{-},\sigma _{k}^{+}\}=1.  \label{CAR}
\end{equation}
The operators commute for different atoms. The Hamiltonian for the beam is
(formally) $H_{R}=\sum_{k=1}^{\infty }\hbar \omega \,\sigma _{k}^{+}\sigma
_{k}^{-}.$

The preparation procedure is the same for each atom and corresponds to an
ensemble state $\varrho $ on $\frak{h}_{R}$. We take the general form 
\begin{equation}
\varrho =p_{1}\sigma ^{+}\sigma ^{-}+p_{0}\sigma ^{-}\sigma ^{+}
\label{state}
\end{equation}
where $p_{1},p_{0}$ are the probabilities to be in the excited state and
ground state respectively. The vacuum state is, of course, specified by $%
p_{0}=1$.

The Hamiltonian is specified by setting 
\begin{equation}
\mathcal{H}_{k}:=\frac{1}{\tau }H_{11}\otimes \sigma _{k}^{+}\sigma _{k}^{-}+%
\frac{1}{\sqrt{\tau }}H_{10}\otimes \sigma _{k}^{+}+\frac{1}{\sqrt{\tau }}%
H_{01}\otimes \sigma _{k}^{-}+H_{00}.
\end{equation}
where we take $H_{11}$ and $H_{00}$ to be self-adjoint and require that $%
\left( H_{01}\right) ^{\dagger }=H_{10}$. We may identify $H_{00}$ with the
free system Hamiltonian $H_{S}$, while $H_{11}$ may be considered to contain
the $H_{R}$ as a component. We shall assume that the operators $H_{\alpha
\beta }$ are bounded with $H_{11}$ also bounded away from zero.

We shall also employ the following summation convention: whenever a repeated
raised and lowered Greek index appears we sum the index over the values zero
and one. With this convention, 
\begin{equation}
\mathcal{H}_{k}\equiv H_{\alpha \beta }\otimes \left[ \frac{\sigma _{k}^{+}}{%
\sqrt{\tau }}\right] ^{\alpha }\left[ \frac{\sigma _{k}^{-}}{\sqrt{\tau }}%
\right] ^{\beta }  \label{Ham}
\end{equation}
were we interpret the raised index as a power: that is, $\left[ x\right]
^{0}=1,\,\left[ x\right] ^{1}=x$.

\subsection{The Collective Operators}

We define the \textit{collective operators} $A^{\pm }\left( t;\tau \right)
,\Lambda \left( t;\tau \right) $ to be 
\begin{gather}
A^{+}\left( t;\tau \right) :=\sqrt{\tau }\sum_{k=1}^{\left\lfloor t/\tau
\right\rfloor }\sigma _{k}^{+};\qquad A^{-}\left( t;\tau \right) :=\sqrt{%
\tau }\sum_{k=1}^{\left\lfloor t/\tau \right\rfloor }\sigma _{k}^{-};  \notag
\\
\Lambda \left( t;\tau \right) :=\sum_{k=1}^{\left\lfloor t/\tau
\right\rfloor }\sigma _{k}^{+}\sigma _{k}^{-}.
\end{gather}
For times $t,s>0$, we have the commutation relations 
\begin{eqnarray*}
\left[ A^{-}\left( t;\tau \right) ,A^{+}\left( s;\tau \right) \right]
&=&\tau \left\lfloor \left( \frac{t\wedge s}{\tau }\right) \right\rfloor
-2\tau \Lambda \left( t\wedge s,\tau \right) , \\
\left[ \Lambda \left( t;\tau \right) ,A^{+}\left( s;\tau \right) \right]
&=&A^{+}\left( t\wedge s;\tau \right) , \\
\left[ A^{-}\left( t;\tau \right) ,\Lambda \left( s;\tau \right) \right]
&=&A^{-}\left( t\wedge s;\tau \right) ,
\end{eqnarray*}
where $s\wedge t$ denotes the minimum of $s$ and $t$. In the limit where $%
\tau $ goes to zero while $s$ and $t$ are held fixed, we have the
approximation 
\begin{equation}
\left[ A^{-}\left( t;\tau \right) ,A^{+}\left( s;\tau \right) \right]
\approx t\wedge s.
\end{equation}
This suggest that the collective fields $A^{\pm }\left( t;\tau \right) $
converge to Bosonic quantum Brownian motions as $\tau \rightarrow 0$ and
that $\Lambda \left( t;\tau \right) $ will converge to the Bosonic
conservation process.

\subsection{Bosonic Noise}

Let $\frak{h}$ be a\ fixed Hilbert space. The $n$-particle Bose states take
the basic form $\phi _{1}\hat{\otimes}\cdots \hat{\otimes}\phi
_{n}=\sum_{\sigma \in \frak{S}_{n}}\phi _{1}\otimes \cdots \otimes \phi _{n}$
where we sum over the permutation group $\frak{S}_{n}$. The $n$-particle
state space is denoted $\frak{h}^{\hat{\otimes}n}$ and the Bose Fock space,
with one particle space $\frak{h}$, is then $\Gamma _{+}\left( \frak{h}%
\right) :=\bigoplus_{n=0}^{\infty }\frak{h}^{\hat{\otimes}n}$ with vacuum
space $\frak{h}^{\hat{\otimes}0}$ spanned by a single vector $\Psi $.

The Bosonic creator, annihilator and differential second quantization fields
are, respectively, the following operators on Fock space 
\begin{eqnarray*}
A^{+}\left( \psi \right) \;\phi _{1}\hat{\otimes}\cdots \hat{\otimes}\phi
_{n} &=&\sqrt{n+1}\,\psi \hat{\otimes}\phi _{1}\hat{\otimes}\cdots \hat{%
\otimes}\phi _{n} \\
A^{-}\left( \psi \right) \;\phi _{1}\hat{\otimes}\cdots \hat{\otimes}\phi
_{n} &=&\frac{1}{\sqrt{n}}\,\sum_{j}\left\langle \psi |\phi \right\rangle 
\hat{\otimes}\phi _{1}\hat{\otimes}\cdots \hat{\otimes}\widehat{\phi _{j}}%
\hat{\otimes}\cdots \hat{\otimes}\phi _{n} \\
d\Gamma \left( T\right) \;\phi _{1}\hat{\otimes}\cdots \hat{\otimes}\phi
_{n} &=&\,\sum_{j}\phi _{1}\hat{\otimes}\cdots \hat{\otimes}\left( T\phi
_{j}\right) \hat{\otimes}\cdots \hat{\otimes}\phi _{n}
\end{eqnarray*}
where $\psi \in \frak{h}$ and $T\in B\left( \frak{h}\right) $.

Now choose $\frak{h}=L^{2}\left( \mathbb{R}^{+},dt\right) $ and set 
\begin{equation}
A_{t}^{\pm }:=A^{\pm }\left( 1_{\left[ 0,t\right] }\right) ;\quad \Lambda
_{t}:=d\Gamma \left( \tilde{1}_{\left[ 0,t\right] }\right)
\end{equation}
where $1_{\left[ 0,t\right] }$ is the characteristic function for the
interval $\left[ 0,t\right] $ and $\tilde{1}_{\left[ 0,t\right] }$ is the
operator on $L^{2}\left( \mathbb{R}^{+},dt\right) $ corresponding to
multiplication by $1_{\left[ 0,t\right] }$.

\bigskip

An integral calculus can be built up around the processes $A_{t}^{\pm
},\Lambda _{t}$ and $t$ and is known as (Bosonic) quantum stochastic
calculus. This allows us to consider quantum stochastic integrals of the
type $\int_{0}^{T}F_{10}\left( t\right) \otimes dA_{t}^{+}+F_{01}\left(
t\right) \otimes dA_{t}^{-}+F_{11}\left( t\right) \otimes d\Lambda
_{t}+F_{00}\left( t\right) \otimes dt$ on $\frak{h}_{0}\otimes \Gamma
_{+}\left( L^{2}\left( \mathbb{R}^{+},dt\right) \right) $ where $\frak{h}%
_{0} $ is some fixed Hilbert space (termed the initial space).

We note the natural isomorphism $\frak{h}_{0}\otimes \Gamma _{+}\left(
L^{2}\left( \mathbb{R}^{+},dt\right) \right) \cong \frak{h}_{t]}\otimes 
\frak{h}_{(t}$ where $\frak{h}_{t]}=\frak{h}_{0}\otimes \Gamma _{+}\left(
L^{2}\left( \left[ 0,t\right] ,dt\right) \right) $ and $\frak{h}_{(t}=\Gamma
_{+}\left( L^{2}\left( (t,\infty ),dt\right) \right) $. A family $\left(
F_{t}\right) _{t}$ of operators on $\frak{h}_{0}\otimes \Gamma _{+}\left(
L^{2}\left( \mathbb{R}^{+},dt\right) \right) $ is said to be adapted if $%
F_{t}$ acts trivially on the future space $\frak{h}_{(t}$\ for each $t$.

The Leibniz rule however breaks down for this theory since products of
stochastic integrals must be put to Wick order before they can be
re-expressed again as stochastic integrals. The new situation is summarized
by the quantum It\^{o} rule $d\left( FG\right) =\left( dF\right) G+F\left(
dG\right) +\left( dF\right) \left( dG\right) $ and the quantum It\^{o} table 
\begin{equation*}
\begin{tabular}{l|llll}
$\times $ & $dA^{+}$ & $d\Lambda $ & $dA^{-}$ & $dt$ \\ \hline
$dA^{+}$ & $0$ & $0$ & $0$ & $0$ \\ 
$d\Lambda $ & $dA^{+}$ & $d\Lambda $ & $0$ & $0$ \\ 
$dA^{-}$ & $dt$ & $dA^{-}$ & $0$ & $0$ \\ 
$dt$ & $0$ & $0$ & $0$ & $0$%
\end{tabular}
\end{equation*}
It is convenient to denote the four basic processes as follows: 
\begin{equation*}
A_{t}^{\alpha \beta }=\left\{ 
\begin{array}{cc}
\Lambda _{t}, & \left( 1,1\right) ; \\ 
A_{t}^{+}, & \left( 1,0\right) ; \\ 
A_{t}^{-}, & \left( 0,1\right) ; \\ 
t, & \left( 0,0\right) .
\end{array}
\right.
\end{equation*}
The It\^{o} table then simplifies to $dA_{t}^{\alpha \beta }dA_{t}^{\mu \nu
}=0$ except for the cases 
\begin{equation}
dA_{t}^{\alpha 1}dA_{t}^{1\beta }=dA_{t}^{\alpha \beta }.  \label{qito}
\end{equation}
The next theorem is from \cite{HP}.

\bigskip

\noindent \textbf{Theorem 2.1: }There exists an unique solution $U_{t}$\ to
the quantum stochastic differential equation (qsde) 
\begin{equation*}
dU_{t}=L_{\alpha \beta }\otimes dA_{t}^{\alpha \beta },\qquad U_{0}=1
\end{equation*}
whenever the coefficients $L_{\alpha \beta }$ are in $B\left( \frak{h}%
_{0}\right) $. The solution is automatically adapted and, moreover, will be
unitary provided that the coefficients take the form 
\begin{equation}
L_{11}=W-1;\quad L_{10}=L;\quad L_{10}=-L^{\dagger }W;\quad L_{00}=-iH-\frac{%
1}{2}L^{\dagger }L  \label{unitarity}
\end{equation}
with $W$ unitary and $H$ self-adjoint.

\subsection{Convergence of the Collective Processes}

Let $\phi ,\psi \in L^{2}\left( \mathbb{R}^{+},dt\right) $ be Riemann
integrable and $T\in L^{\infty }\left( \mathbb{R}^{+},dt\right) $
continuous. We define the pre-limit fields 
\begin{gather*}
A^{+}\left( \phi ,\tau \right) :=\sqrt{\tau }\sum_{k}\phi \left( \tau
k\right) \,\sigma _{k}^{+},\quad A^{-}\left( \psi ,\tau \right) :=\sqrt{\tau 
}\sum_{k}\psi ^{\ast }\left( \tau k\right) \,\sigma _{k}^{-} \\
\Lambda \left( T;\tau \right) :=\sum_{k=1}^{\left\lfloor t/\tau
\right\rfloor }T\left( \tau k\right) \,\sigma _{k}^{+}\sigma _{k}^{-}.
\end{gather*}
then 
\begin{equation*}
\left[ A^{-}\left( \psi ,\tau \right) ,A^{+}\left( \phi ,\tau \right) \right]
=\tau \sum_{k}\psi ^{\ast }\left( \tau k\right) \phi \left( \tau k\right)
-2\tau \sum_{k}\psi ^{\ast }\left( \tau k\right) \phi \left( \tau k\right)
\,\sigma _{k}^{+}\sigma _{k}^{-}.
\end{equation*}
which converges to $\left\langle \psi |\phi \right\rangle =\int_{0}^{\infty
}\psi ^{\ast }\left( t\right) \phi \left( t\right) dt$ as $\tau \rightarrow
0 $ in the vacuum state $\Phi ^{\tau }$.

More generally we have convergence of the type 
\begin{eqnarray}
&&\lim_{\tau \rightarrow 0}\left\langle \Phi ^{\tau }|\,\prod_{j=1,\cdots
,n}^{\rightarrow }\exp \left\{ i\Lambda \left( T_{j},\tau \right)
+iA^{+}\left( \phi _{j},\tau \right) +iA^{-}\left( \psi _{j},\tau \right)
\right\} \,\Phi ^{\tau }\right\rangle  \notag \\
&=&\left\langle \Psi |\,\prod_{j=1,\cdots ,n}^{\rightarrow }\exp \left\{
i\Lambda \left( T_{j}\right) +iA^{+}\left( \phi _{j}\right) +iA^{-}\left(
\psi _{j}\right) \right\} \,\Psi \right\rangle
\end{eqnarray}
where $\overrightarrow{\prod }_{j=1,...,n}X_{j}=X_{1}X_{2}\cdots X_{n}$
denotes an ordered product of operators.

\section{Decomposition of the Floquet operators}

Let $\sigma ^{\pm }$ be the two-level transition operators. We set 
\begin{equation*}
\mathcal{H}=H_{\alpha \beta }\otimes \left[ \frac{\sigma ^{+}}{\sqrt{\tau }}%
\right] ^{\alpha }\left[ \frac{\sigma ^{-}}{\sqrt{\tau }}\right] ^{\beta }
\end{equation*}
and assume that $\left\| H_{\alpha \beta }\right\| \leq C$. Using the
anti-commutation relations, we have that 
\begin{eqnarray*}
\left( \tau \mathcal{H}\right) ^{2} &=&\left[ H_{11}H_{11}+\tau
H_{11}H_{00}+\tau H_{10}H_{01}-\tau H_{01}H_{10}+H_{00}H_{11}\right] \sigma
^{+}\sigma ^{-} \\
&&+\sqrt{\tau }\left[ H_{11}H_{10}+\tau H_{10}H_{00}+\tau H_{00}H_{10}\right]
\sigma ^{+} \\
&&+\sqrt{\tau }\left[ H_{01}H_{11}+\tau H_{01}H_{00}+\tau H_{00}H_{01}\right]
\sigma ^{-} \\
&&+\tau \left[ H_{01}H_{10}+\tau H_{00}H_{00}\right] .
\end{eqnarray*}
More generally, for $m\geq 2$, we have that (recall our summation
convention!) 
\begin{equation}
\left( \tau \mathcal{H}\right) ^{n}=\tau ^{n}H_{\alpha _{n}\beta _{n}}\cdots
H_{\alpha _{1}\beta _{1}}\otimes \left[ \frac{\sigma ^{+}}{\sqrt{\tau }}%
\right] ^{\alpha _{n}}\left[ \frac{\sigma ^{-}}{\sqrt{\tau }}\right] ^{\beta
_{n}}\cdots \left[ \frac{\sigma ^{+}}{\sqrt{\tau }}\right] ^{\alpha _{1}}%
\left[ \frac{\sigma ^{-}}{\sqrt{\tau }}\right] ^{\beta _{1}}
\label{nth term}
\end{equation}
and, again by repeated use of the anti-commutation relations, we are lead to
the form 
\begin{equation}
\left( \tau \mathcal{H}\right) ^{n}=\tau \,K_{\alpha \beta }^{\left(
n\right) }\left( \tau \right) \otimes \left[ \frac{\sigma ^{+}}{\sqrt{\tau }}%
\right] ^{\alpha }\left[ \frac{\sigma ^{-}}{\sqrt{\tau }}\right] ^{\beta }
\end{equation}
where 
\begin{equation}
K_{\alpha \beta }^{\left( n\right) }\left( \tau \right) :=H_{\alpha 1}\left(
H_{11}\right) ^{m-2}H_{1\beta }+O\left( \tau \right) .
\end{equation}
The remainder is a polynomial of degree $m$ in $\tau $, whose coefficients
are sums of $n$-fold products of the $H_{\alpha \beta }$'s, having no
constant term. Here $O\left( \tau \right) $ means a term going to zero in
operator-norm faster than $\tau $ as $\tau \rightarrow 0$.

We next of all compute the Floquet operator: 
\begin{equation}
V=\exp \left\{ -i\tau \mathcal{H}\right\} =1+\tau \left\{ L_{\alpha \beta
}+R_{\alpha \beta }\left( \tau \right) \right\} \otimes \left[ \frac{\sigma
^{+}}{\sqrt{\tau }}\right] ^{\alpha }\left[ \frac{\sigma ^{-}}{\sqrt{\tau }}%
\right] ^{\beta }.  \label{series}
\end{equation}
Here the $L_{\alpha \beta }$ and $R_{\alpha \beta }\left( \tau \right) $ are
bounded operators on the system space with $R_{\alpha \beta }\left( \tau
\right) =O\left( \tau \right) $. Explicitly, the coefficients $L_{\alpha
\beta }$ are given by 
\begin{eqnarray}
L_{11} &=&e^{-iH_{11}}-1;  \notag \\
L_{10} &=&\frac{e^{-iH_{11}}-1}{H_{11}}H_{10};  \notag \\
L_{01} &=&H_{01}\frac{e^{-iH_{11}}-1}{H_{11}};  \notag \\
L_{00} &=&-iH_{00}+H_{01}\frac{e^{-iH_{11}}-1+iH_{11}}{\left( H_{11}\right)
^{2}}H_{10}.  \label{ItoHolevo}
\end{eqnarray}
We remark that these coefficients take the form (\ref{unitarity})\ where $%
W:=\exp \left\{ -iH_{11}\right\} $ is unitary, $H:=H_{00}-H_{01}\frac{%
H_{11}-\sin \left( H_{11}\right) }{\left( H_{11}\right) ^{2}}H_{10}$ is
self-adjoint and $L$ is bounded but otherwise arbitrary. (Note that $\frac{%
x-\sin x}{x^{2}}>0$ for $x>0$.)

\section{Limit For the Vacuum State}

We begin by assuming that $p_{1}=0$ and that therefore the reservoir is in
the vacuum state $\Phi ^{\tau }$.

Let $t>0$, then we are interested in the unitary $U_{t}^{\left( \tau \right)
}=V_{\left\lfloor t/\tau \right\rfloor }\cdots V_{2}V_{1}$ in the limit $%
\tau \rightarrow 0$. By virtue of an uniform estimate established in the
next section, we have that the components $R_{\alpha \beta }\left( \tau
\right) $ make negligible contribution in the limit $\tau \rightarrow 0$. It
is easy to see that, if $k=\left\lfloor t/\tau \right\rfloor $ and if we
ignore the negligible component,then 
\begin{eqnarray*}
\frac{1}{\tau }\left( U_{t+\tau }^{\left( \tau \right) }-U_{t}^{\left( \tau
\right) }\right) &=&\frac{1}{\tau }\left( V_{k+1}-1\right) U_{t}^{\left(
\tau \right) } \\
&=&L_{\alpha \beta }\otimes \left[ \frac{\sigma _{k+1}^{+}}{\sqrt{\tau }}%
\right] ^{\alpha }\left[ \frac{\sigma _{k+1}^{-}}{\sqrt{\tau }}\right]
^{\beta }U_{t}^{\left( \tau \right) }+\cdots .
\end{eqnarray*}
We shall replace this by a quantum stochastic differential equation shortly.
The operator $U_{t}^{\left( \tau \right) }$ is then represented as 
\begin{equation}
U_{t}^{\left( \tau \right) }=\prod_{j=\left\lfloor t/\tau \right\rfloor
,\cdots ,1}^{\rightarrow }\left\{ 1+\tau \left[ L_{\alpha \beta }+R_{\alpha
\beta }\left( \tau \right) \right] \otimes \left[ \frac{\sigma _{j}^{+}}{%
\sqrt{\tau }}\right] ^{\alpha \left( j\right) }\left[ \frac{\sigma _{j}^{-}}{%
\sqrt{\tau }}\right] ^{\beta \left( j\right) }\right\}  \label{aa}
\end{equation}

\bigskip

\noindent \textbf{Theorem 4.1: }In the above notations, the discrete time
family $\left\{ U_{t}^{\left( \tau \right) }\right\} $ converges to quantum
stochastic process $U_{t}$\ on $\frak{h}_{S}\otimes \Gamma _{+}\left(
L^{2}\left( \mathbb{R+},dt\right) \right) $ in the sense that, for all $%
u,v\in \frak{h}_{S}$, integers $n,m$ and for all $\phi _{j},\psi _{j}\in
L^{2}\left( \mathbb{R}^{+},dt\right) $ Riemann integrable, we have the
uniform convergence 
\begin{gather}
\left\langle A^{+}\left( \phi _{m},\tau \right) \cdots A^{+}\left( \phi
_{1},\tau \right) \,u\otimes \Phi ^{\tau }|\,U_{t}^{\left( \tau \right)
}\,A^{+}\left( \psi _{n},\tau \right) \cdots A^{+}\left( \psi _{1},\tau
\right) \,v\otimes \Phi ^{\tau }\right\rangle  \notag \\
\rightarrow \left\langle A^{+}\left( \phi _{m}\right) \cdots A^{+}\left(
\phi _{1}\right) \,u\otimes \Psi |\,U_{t}\,A^{+}\left( \psi _{n}\right)
\cdots A^{+}\left( \psi _{1}\right) \,v\otimes \Psi \right\rangle
\end{gather}
The process $U_{t}$ is moreover unitary, adapted and satisfies the (quantum)
stochastic differential equation 
\begin{equation}
dU_{t}=L_{\alpha \beta }\otimes dA_{t}^{\alpha \beta }\,U_{t},\quad U_{0}=1.
\label{qsde}
\end{equation}

\bigskip

\noindent \textbf{Remark 1: }The solution to (\ref{qsde}) can be written as 
\begin{equation*}
U_{t}=\mathbf{\vec{N}}\exp \left\{ \int_{0}^{t}ds\,L_{\alpha \beta }\left[
a_{s}^{+}\right] ^{\alpha }\left[ a_{s}^{-}\right] ^{\beta }\right\}
\end{equation*}
where $a_{t}^{\pm }$ are quantum white noises. Here the symbol $\mathbf{\vec{%
N}}$ stands for normal ordering and we understand the formal development 
\begin{eqnarray*}
U_{t} &=&1+\sum_{n=1}^{\infty }\int_{t>t_{1}>\dots t_{n}>0}dt_{n}\dots
dt_{1}\,L_{\alpha _{n}\beta _{n}}\cdots L_{\alpha _{1}\beta _{1}} \\
&&\times \left[ a_{t_{n}}^{+}\right] ^{\alpha _{n}}\cdots \left[
a_{t_{1}}^{+}\right] ^{\alpha _{1}}\left[ a_{t_{1}}^{-}\right] ^{\beta
_{1}}\cdots \left[ a_{t_{n}}^{-}\right] ^{\beta _{n}}.
\end{eqnarray*}

The connection with the Hudson-Parthasarathy notation is made by the
replacements $\left[ a_{t}^{+}\right] ^{\alpha }K_{t}\left[ a_{t}^{-}\right]
^{\beta }dt\hookrightarrow K_{t}dA_{t}^{\alpha \beta }$ for adapted $K_{t}$.

\bigskip

\noindent \textbf{Remark 2: }There is an alternative presentation \cite
{Holevo}\ which we refer to as the Holevo ordered form. We write 
\begin{equation*}
U_{t}=\mathbf{\vec{H}}\exp \left\{ \int_{0}^{t}ds\,H_{\alpha \beta }\left[
a_{s}^{+}\right] ^{\alpha }\left[ a_{s}^{-}\right] ^{\beta }\right\}
\end{equation*}
and understand this to be the It\^{o} qsde 
\begin{equation*}
dU_{t}=\left( e^{H_{\alpha \beta }dA_{t}^{\alpha \beta }}-1\right)
U_{t};\qquad U_{0}=1.
\end{equation*}
From the quantum It\^{o} table we have that

\begin{eqnarray*}
e^{H_{\alpha \beta }dA_{t}^{\alpha \beta }}-1 &=&\sum_{n=1}^{\infty }\frac{%
\left( -i\right) ^{n}}{n!}H_{\alpha _{n}\beta _{n}}\dots H_{\alpha _{i}\beta
_{1}}\;dA_{t}^{\alpha _{n}\beta _{n}}\dots dA_{t}^{\alpha _{1}\beta _{1}} \\
&=&H_{\alpha \beta }\;dA_{t}^{\alpha \beta }+\sum_{n=2}^{\infty }\frac{%
\left( -i\right) ^{n}}{n!}H_{\alpha 1}\left( H_{11}\right) ^{n-2}H_{1\beta
}\;dA_{t}^{\alpha \beta }.
\end{eqnarray*}
We see that the It\^{o} coefficients, $L_{\alpha \beta }$, and Holevo
coefficients, $H_{\alpha \beta }$, are connected according to the same
relations as (\ref{ItoHolevo}).

\bigskip

\noindent \textbf{Remark 3: }The theorem may be restated in a more elegant
fashion. The discrete unitary process $U_{t}\left( \tau \right) =\mathbf{%
\vec{T}}\exp \left\{ -i\int_{0}^{t}H_{s}\left( \tau \right) ds\right\} $,
where $H_{t}\left( \tau \right) =\sum_{k}\delta \left( t-k\tau \right) 
\mathcal{H}_{k}$ with $\mathcal{H}_{k}$ given by (\ref{Ham}), converges to
the continuous-time unitary process $U_{t}=\mathbf{\vec{H}}\exp \left\{
\int_{0}^{t}ds\,H_{\alpha \beta }\left[ a_{s}^{+}\right] ^{\alpha }\left[
a_{s}^{-}\right] ^{\beta }\right\} $.

\bigskip

\noindent \textbf{Remark 4: }The basic estimates in section 4 of \cite
{Holevo} serve to show the convergence of (\ref{aa}) to the Holevo ordered
form.

\bigskip

Next of all we turn our attention to the Heisenberg evolution. We begin by
noting that if we set $\Delta A^{\alpha \beta }=\tau \left[ \frac{\sigma ^{+}%
}{\sqrt{\tau }}\right] ^{\alpha }\left[ \frac{\sigma ^{-}}{\sqrt{\tau }}%
\right] ^{\beta }$ then 
\begin{equation*}
\Delta A^{\alpha 1}\Delta A^{1\beta }=\tau \left[ \frac{\sigma ^{+}}{\sqrt{%
\tau }}\right] ^{\alpha }\left[ \frac{\sigma ^{-}}{\sqrt{\tau }}\right]
^{1}\times \tau \left[ \frac{\sigma ^{+}}{\sqrt{\tau }}\right] ^{1}\left[ 
\frac{\sigma ^{-}}{\sqrt{\tau }}\right] ^{\beta }=\Delta A^{\alpha \beta }
\end{equation*}
and that otherwise $\Delta A^{\alpha \beta }\Delta A^{\mu \nu }=O\left( \tau
\right) \Delta A^{\eta \xi }$. This is the discrete form of the quantum
It\^{o} table (\ref{qito}).

Let $J_{t}^{\left( \tau \right) }\left( X\right) =U_{t}^{\left( \tau \right)
\dagger }\left( X\otimes 1\right) U_{t}^{\left( \tau \right) }$. If $%
k=\left\lfloor t/\tau \right\rfloor ,$ then 
\begin{gather*}
\frac{1}{\tau }\left( J_{t+\tau }^{\left( \tau \right) }\left( X\right)
-J_{t}^{\left( \tau \right) }\left( X\right) \right) = \\
\left\{ J_{t}^{\left( \tau \right) }\left( L_{\beta \alpha }^{\dagger
}X\right) +J_{t}^{\left( \tau \right) }\left( XL_{\alpha \beta }\right)
+J_{t}^{\left( \tau \right) }\left( L_{1\alpha }^{\dagger }XL_{1\beta
}\right) +S_{\alpha \beta }\left( \tau \right) \right\} \otimes \left[ \frac{%
\sigma _{k+1}^{+}}{\sqrt{\tau }}\right] ^{\alpha }\left[ \frac{\sigma
_{k+1}^{-}}{\sqrt{\tau }}\right] ^{\beta }
\end{gather*}
where $S_{\alpha \beta }\left( \tau \right) =O\left( \tau \right) $. Again
the terms $S_{\alpha \beta }\left( \tau \right) $ will have negligible
contribution in the $\tau \rightarrow 0$ limit. We establish the appropriate
uniform estimate in the next section.

\bigskip

\noindent \textbf{Theorem 4.2: }In the above notations, the discrete time
family $\left\{ J_{t}^{\left( \tau \right) }\left( X\right) \right\} $
converges to quantum stochastic process $J_{t}\left( X\right)
=U_{t}^{\dagger }\left( X\otimes 1\right) U_{t}$\ on $\frak{h}_{S}\otimes
\Gamma _{+}\left( L^{2}\left( \mathbb{R+},dt\right) \right) $ in the sense
that, for all $u,v\in \frak{h}_{S}$, integers $n,m$ and for all $\phi
_{j},\psi _{j}\in L^{2}\left( \mathbb{R}^{+},dt\right) $ Riemann integrable,
we have the uniform convergence 
\begin{gather}
\left\langle A^{+}\left( \phi _{m},\tau \right) \cdots A^{+}\left( \phi
_{1},\tau \right) \,u\otimes \Phi ^{\tau }|\,J_{t}^{\left( \tau \right)
}\left( X\right) \,A^{+}\left( \psi _{n},\tau \right) \cdots A^{+}\left(
\psi _{1},\tau \right) \,v\otimes \Phi ^{\tau }\right\rangle  \notag \\
\rightarrow \left\langle A^{+}\left( \phi _{m}\right) \cdots A^{+}\left(
\phi _{1}\right) \,u\otimes \Psi |\,J_{t}\left( X\right) \,A^{+}\left( \psi
_{n}\right) \cdots A^{+}\left( \psi _{1}\right) \,v\otimes \Psi
\right\rangle .
\end{gather}
The process $U_{t}$ is moreover unitary, adapted and satisfies the (quantum)
stochastic differential equation 
\begin{equation}
dJ_{t}\left( X\right) =J_{t}\left( \mathcal{L}_{\alpha \beta }X\right)
\otimes dA_{t}^{\alpha \beta },\quad J_{0}\left( X\right) =X\otimes 1
\end{equation}
where 
\begin{equation}
\mathcal{L}_{\alpha \beta }X:=L_{\beta \alpha }^{\dagger }X+XL_{\alpha \beta
}+L_{1\alpha }^{\dagger }XL_{1\beta }.
\end{equation}

\bigskip

\noindent\ \textbf{Remark 5}\ A completely positive semigroup $\left\{ \Xi
_{t}:t\geq 0\right\} $ is then defined on $B\left( \frak{h}_{S}\right) $ by $%
\left\langle u\otimes \Psi |\,J_{t}\left( X\right) \,v\otimes \Psi
\right\rangle :=\left\langle u|\,\Xi _{t}\left( X\right) \,v\right\rangle $
and we have $\Xi _{t}=\exp \left\{ t\mathcal{L}_{00}\right\} $ where the
Lindblad generator is 
\begin{equation*}
\mathcal{L}_{00}\left( X\right) =\frac{1}{2}\left[ L^{\dagger },X\right] L+%
\frac{1}{2}L^{\dagger }\left[ X,L\right] -i\left[ X,H\right]
\end{equation*}
where $L=\frac{e^{-iH_{11}}-1}{H_{11}}H_{10}$ and $H=H_{00}-H_{01}\frac{%
H_{11}-\sin \left( H_{11}\right) }{\left( H_{11}\right) ^{2}}H_{10}$.

\section{Uniform Estimates}

We now want to obtain a norm-estimate for the series (\ref{series}) based on
the expansion appearing in (\ref{nth term}).

\bigskip

\noindent \textbf{Lemma 5.1:} In the notations of the previous section 
\begin{gather*}
\sum_{n\geq 0}\frac{\tau ^{n}}{n!}\sum_{\alpha ,\beta \in \left\{
0,1\right\} ^{n}}\left\| H_{\alpha _{n}\beta _{n}}\cdots H_{\alpha _{1}\beta
_{1}}\otimes \left[ \frac{\sigma ^{+}}{\sqrt{\tau }}\right] ^{\alpha _{n}}%
\left[ \frac{\sigma ^{-}}{\sqrt{\tau }}\right] ^{\beta _{n}}\cdots \left[ 
\frac{\sigma ^{+}}{\sqrt{\tau }}\right] ^{\alpha _{1}}\left[ \frac{\sigma
^{-}}{\sqrt{\tau }}\right] ^{\beta _{1}}\right\| \\
\leq \exp \left\{ \tau \left( e^{C}-1\right) \right\} .
\end{gather*}
(Recall that $C=\max_{\alpha \beta }\left\| H_{\alpha \beta }\right\| ,$ so $%
e^{C}>1$, and note that we have broken from our summation convention to show
explicitly that we have a sum of norms.)

\begin{proof}
Evidently we have the bound 
\begin{eqnarray}
&&\tau ^{n}\sum_{\alpha ,\beta \in \left\{ 0,1\right\} ^{n}}\left\|
H_{\alpha _{n}\beta _{n}}\cdots H_{\alpha _{1}\beta _{1}}\otimes \left[ 
\frac{\sigma ^{+}}{\sqrt{\tau }}\right] ^{\alpha _{n}}\left[ \frac{\sigma
^{-}}{\sqrt{\tau }}\right] ^{\beta _{n}}\cdots \left[ \frac{\sigma ^{+}}{%
\sqrt{\tau }}\right] ^{\alpha _{1}}\left[ \frac{\sigma ^{-}}{\sqrt{\tau }}%
\right] ^{\beta _{1}}\right\|  \notag \\
&\leq &\tau ^{n}C^{n}\sum_{\alpha ,\beta \in \left\{ 0,1\right\}
^{n}}\left\| \left[ \frac{\sigma ^{+}}{\sqrt{\tau }}\right] ^{\alpha _{n}}%
\left[ \frac{\sigma ^{-}}{\sqrt{\tau }}\right] ^{\beta _{n}}\cdots \left[ 
\frac{\sigma ^{+}}{\sqrt{\tau }}\right] ^{\alpha _{1}}\left[ \frac{\sigma
^{-}}{\sqrt{\tau }}\right] ^{\beta _{1}}\right\| .  \label{tobebound}
\end{eqnarray}
Now $\tau ^{n}\left[ \frac{\sigma ^{+}}{\sqrt{\tau }}\right] ^{\alpha _{n}}%
\left[ \frac{\sigma ^{-}}{\sqrt{\tau }}\right] ^{\beta _{n}}\cdots \left[ 
\frac{\sigma ^{+}}{\sqrt{\tau }}\right] ^{\alpha _{1}}\left[ \frac{\sigma
^{-}}{\sqrt{\tau }}\right] ^{\beta _{1}}$ can be described as follows: we
have $n$ ordered vertices labelled $j=1,\cdots ,n$ and at the $k$-th vertex
will be either $\sigma ^{+}\sigma ^{-},\sqrt{\tau }\sigma ^{+},\sqrt{\tau }%
\sigma ^{-}$\ or $\tau $ depending on whether $\left( \alpha j,\beta
_{j}\right) =\left( 1,1\right) ,\left( 1,0\right) ,\left( 0,1\right) $ or $%
\left( 0,0\right) $ respectively. Our first objective is to put this
expression to Wick order. In placing the $\sigma ^{+}$'s to the left of the $%
\sigma ^{-}$'s, we must repeatedly use the anti-commutation relations. This
introduces, in the usual way, the notion of pair contractions between a $%
\sigma ^{+}$ and a $\sigma ^{-}$ at different vertices- that is, we replace $%
A\sigma ^{+}B\sigma ^{-}C$ with $ABC$. (We ignore the possible minus sign
occurring as we want a norm estimate.)

We then have to contend with a sum over all possible pair contractions. We
have at most one creator and one annihilator at each vertex. Therefore, in a
typical term we shall have several vertices connected through pair
contractions and these vertices form disjoint subsets of all the $n$
vertices. We also take each $\left( 0,0\right) $ vertex to be a singleton
set. In this way each term corresponds to a partition on the $n$ vertices
into subsets. Now recall that the number of ways to partition $n$ objects
into $m$ subsets is given by Stirling's number, $S\left( n,m\right) $, of
the second kind \cite{Riodain}. Each subset of the partition contributes a
factor $\tau $: this is obvious for the singletons and more generally we
have a subset with $\sqrt{\tau }$ for both of the terminal vertices and
unity for the internal scattering vertices. We have in addition a product of
the uncontracted $\sigma ^{\pm }$ but this will have norm bounded by unity.

This leads to the following bound for (\ref{tobebound}): $%
\sum_{m=1}^{n}C^{n}\tau ^{m}S\left( n,m\right) .$The norm-bound for the
expansion for the full series is then 
\begin{equation}
\sum_{n=0}^{\infty }\frac{1}{n!}\sum_{m=1}^{n}C^{n}\tau ^{m}S\left(
n,m\right) =\exp \left\{ \tau \left( e^{C}-1\right) \right\}  \label{bound}
\end{equation}
where we use the well known generating series \cite{Riodain} for the
Stirling numbers.
\end{proof}

(We remark that the integer $S\left( n,m\right) $ gives the coefficient of $%
\lambda ^{m}$ in the $n$-th moment of a Poisson distributed random variable
with intensity $\lambda $.)

\bigskip

\noindent \textbf{Corollary 5.2:} 
\begin{gather*}
\sum_{k=1}^{\left\lfloor t/\tau \right\rfloor }\sum_{n_{1},\cdots ,n_{k}\geq
0}\frac{\tau ^{n_{1}+\cdots +n_{k}}}{n_{1}!\cdots n_{k}!} \\
\times \prod_{j}\sum_{\alpha ,\beta \in \left\{ 0,1\right\} ^{n_{j}}}\left\|
H_{\alpha _{n_{j}}\beta _{n_{j}}}\cdots H_{\alpha _{1}\beta _{1}}\otimes 
\left[ \frac{\sigma ^{+}}{\sqrt{\tau }}\right] ^{\alpha _{nj}}\left[ \frac{%
\sigma ^{-}}{\sqrt{\tau }}\right] ^{\beta _{n_{j}}}\cdots \left[ \frac{%
\sigma ^{+}}{\sqrt{\tau }}\right] ^{\alpha _{1}}\left[ \frac{\sigma ^{-}}{%
\sqrt{\tau }}\right] ^{\beta _{1}}\right\| \\
\leq \exp \left\{ t\left( e^{C}-1\right) \right\} .
\end{gather*}

\bigskip

This establishes a uniform estimate for the series expansion of $%
U_{t}=V_{\left\lfloor t/\tau \right\rfloor }\cdots V_{2}V_{1}$ based on the
development (\ref{series}). We now do the same for the Heisenberg evolution.
For $X\in B\left( \frak{h}_{S}\right) $ we have 
\begin{gather*}
V^{\dagger }\left( X\otimes 1\right) V=\sum_{n,n^{\prime }}\left( -i\right)
^{n^{\prime }-n}\frac{\tau ^{n+n^{\prime }}}{n!n^{\prime }!}H_{\alpha
_{n}\beta _{n}}\cdots H_{\alpha _{1}\beta _{1}}\,X\,H_{\mu _{n^{\prime }}\nu
_{n^{\prime }}}\cdots H_{\mu _{1}\nu _{1}} \\
\otimes \left[ \frac{\sigma ^{+}}{\sqrt{\tau }}\right] ^{\alpha _{n}}\left[ 
\frac{\sigma ^{-}}{\sqrt{\tau }}\right] ^{\beta _{n}}\cdots \left[ \frac{%
\sigma ^{+}}{\sqrt{\tau }}\right] ^{\alpha _{1}}\left[ \frac{\sigma ^{-}}{%
\sqrt{\tau }}\right] ^{\beta _{1}}\times \left[ \frac{\sigma ^{+}}{\sqrt{%
\tau }}\right] ^{\mu _{n^{\prime }}}\left[ \frac{\sigma ^{-}}{\sqrt{\tau }}%
\right] ^{\nu _{n^{\prime }}}\cdots \left[ \frac{\sigma ^{+}}{\sqrt{\tau }}%
\right] ^{\mu _{1}}\left[ \frac{\sigma ^{-}}{\sqrt{\tau }}\right] ^{\nu _{1}}
\end{gather*}
and by the previous arguments we see that the sum of the norms of these
summands can be bounded by 
\begin{equation*}
\left\| X\right\| \sum_{n,n^{\prime }}\sum_{m,m^{\prime }}\frac{%
C^{n+n^{\prime }}\tau ^{m+m^{\prime }}}{n!n^{\prime }!}S\left( n,m\right)
S\left( n^{\prime },m^{\prime }\right) =\left\| X\right\| \exp \left\{ 2\tau
\left( e^{C}-1\right) \right\} .
\end{equation*}
Likewise, the series expansion of $J_{t}^{\left( \tau \right) }\left(
X\right) =U_{t}^{\left( \tau \right) \dagger }\left( X\otimes 1\right)
U_{t}^{\left( \tau \right) }$ in terms of the fundamental Hamiltonian
components will be bounded by $\left\| X\right\| \exp \left\{ 2t\left(
e^{C}-1\right) \right\} $ uniformly.

\section{Non-Vacuum State}

\subsection{Gaussian Case}

We consider the situation where $H_{11}=0$. Let $\left\langle .\right\rangle 
$ be the state determined by the density matrix $\varrho $ in (\ref{state}).
With the convention that $\sigma ^{z}=2\sigma ^{+}\sigma ^{-}-1$, we set 
\begin{equation*}
\hat{\sigma}_{k}^{\pm }=\sqrt{p_{0}}\,\sigma _{k}^{\pm }\otimes 1+\sqrt{p_{1}%
}\,\sigma _{k}^{z}\otimes \sigma _{k}^{\mp }
\end{equation*}
These operators commute for different lables $k$ and we have $\left\{ \hat{%
\sigma}_{k}^{+},\hat{\sigma}_{k}^{-}\right\} =1$, $\left( \hat{\sigma}%
_{k}^{\pm }\right) ^{2}=0$. Moreover, algebra generated by the $\hat{\sigma}%
_{k}^{\pm }$ with the pure state $e_{0}\otimes e_{0}$ is isomorphic to the
one generated by the $\sigma _{k}^{\pm }$ with mixed state $\left\langle
.\right\rangle $.

If we adopt the new representation then we can consider collective fields 
\begin{equation*}
\hat{A}^{\pm }\left( \phi ,\tau \right) =\sqrt{p_{0}}\,B^{\pm }\left( \phi
,\tau \right) +\sqrt{p_{1}}\,C^{\mp }\left( j\phi ,\tau \right)
\end{equation*}
where 
\begin{eqnarray*}
B^{+}\left( \phi ,\tau \right) &:&=\sqrt{\tau }\sum_{k}\phi \left( k\tau
\right) \,\sigma _{k}^{+}\otimes 1; \\
C^{-}\left( \phi ,\tau \right) &:&=\sqrt{\tau }\sum_{k}\phi \left( k\tau
\right) \,\sigma _{k}^{z}\otimes \sigma _{k}^{-}.
\end{eqnarray*}
Here $j:\phi \mapsto \phi ^{\ast }$ is the complex conjugation. We note that 
\begin{equation*}
\left[ B^{-}\left( \phi ,\tau \right) ,C^{+}\left( \psi ,\tau \right) \right]
=2\tau \sum_{k}\phi \left( k\tau \right) ^{\ast }\psi \left( k\tau \right)
\,\sigma _{k}^{+}\otimes \sigma _{k}^{-}
\end{equation*}
and that this and similar terms are negligible in the limit $\tau
\rightarrow 0$ for the state $\Phi ^{\tau }\otimes \Phi ^{\tau }$.

We therefore find that the fields $B^{\sharp }\left( .,\tau \right) $ and $%
C^{\sharp }\left( .,\tau \right) $ in the state $\Phi ^{\tau }\otimes \Phi
^{\tau }$ converge in distribution to independent (that is, commuting) Bose
fields $B^{\sharp }\left( .\right) $ and $C^{\sharp }\left( .\right) $ with\
the double Fock vacuum state $\Psi \otimes \Psi $. We set $\hat{A}\left(
.\right) =\sqrt{p_{0}}\,B^{\pm }\left( .\right) +\sqrt{p_{1}}\,C^{\mp
}\left( j.\right) $ and find the modified It\^{o} table 
\begin{equation*}
d\hat{A}_{t}^{-}d\hat{A}_{t}^{+}=p_{0}\,dB_{t}^{-}dB_{t}^{+}=p_{0}\,dt,%
\qquad d\hat{A}_{t}^{+}d\hat{A}_{t}^{-}=p_{1}\,dC_{t}^{-}dC_{t}^{+}=p_{1}%
\,dt.
\end{equation*}
It is relatively easy to see that there again exists\ a limit process $U_{t}$%
, this time driven by the fields $\hat{A}_{t}^{\pm }$. The limit qsde will
be 
\begin{equation}
dU_{t}=\left\{ H_{10}\otimes d\hat{A}_{t}^{+}+H_{01}\otimes d\hat{A}%
_{t}^{-}-\left( iH_{00}+\frac{1}{2}p_{0}H_{01}H_{10}+\frac{1}{2}%
p_{1}H_{10}H_{01}\right) \right\} U_{t}
\end{equation}
which is again unitary and adapted to the noise fields $\hat{A}_{t}^{\pm }$.

\subsection{Asymptotic Case}

We consider a new orthonormal basis for $\frak{h}_{R}$ 
\begin{eqnarray*}
\left| \tilde{e}_{0}\right\rangle &=&\theta \sqrt{p_{0}}\left|
e_{0}\right\rangle +\theta ^{\ast }\sqrt{p_{1}}\left| e_{1}\right\rangle ; \\
\left| \tilde{e}_{1}\right\rangle &=&-\theta ^{\ast }\sqrt{p_{1}}\left|
e_{0}\right\rangle +\theta \sqrt{p_{0}}\left| e_{1}\right\rangle .
\end{eqnarray*}
where $\theta $ is a complex number of unit modulus and $p_{0}+p_{1}=0$.
Introduce new transition operators 
\begin{equation*}
\tilde{\sigma}^{+}=\left| \tilde{e}_{1}\right\rangle \left\langle \tilde{e}%
_{0}\right| ,\tilde{\sigma}^{-}=\left| \tilde{e}_{0}\right\rangle
\left\langle \tilde{e}_{1}\right|
\end{equation*}
The new variables $\tilde{\sigma}^{\pm }$ again satisfy the proper
anti-commutation relations however $\tilde{\sigma}^{-}$ annihilates the
state $\tilde{e}_{0}$.

We note that $\left\langle \tilde{e}_{0}\right| \sigma ^{+}\sigma ^{-}\left| 
\tilde{e}_{0}\right\rangle =p_{1},$ $\left\langle \tilde{e}_{0}\right|
\sigma ^{-}\sigma ^{+}\left| \tilde{e}_{0}\right\rangle =p_{0},$ $%
\left\langle \tilde{e}_{0}\right| \sigma ^{+}\left| \tilde{e}%
_{0}\right\rangle =\sqrt{p_{1}p_{0}}\theta ^{2}$ and $\left\langle \tilde{e}%
_{0}\right| \sigma ^{-}\left| \tilde{e}_{0}\right\rangle =\sqrt{p_{1}p_{0}}%
\theta ^{\ast 2}$. We find that 
\begin{eqnarray*}
\sigma ^{+} &=&p_{0}\tilde{\sigma}^{+}-p_{1}\tilde{\sigma}^{-}-\sqrt{%
p_{1}p_{0}}\left( \theta ^{2}+\theta ^{\ast 2}\right) \tilde{\sigma}^{+}%
\tilde{\sigma}^{-}+\sqrt{p_{1}p_{0}}\theta ^{2}; \\
\sigma ^{+}\sigma ^{-} &=&\left( p_{0}-p_{1}\right) \tilde{\sigma}^{+}\tilde{%
\sigma}^{-}-\sqrt{p_{1}p_{0}}\left( \theta ^{2}+\theta ^{\ast 2}\right)
\left( \tilde{\sigma}^{+}+\tilde{\sigma}^{-}\right) +p_{1}.
\end{eqnarray*}
After some rearrangement, we obtain 
\begin{equation*}
H_{\alpha \beta }\otimes \left[ \frac{\sigma ^{+}}{\sqrt{\tau }}\right]
^{\alpha }\left[ \frac{\sigma ^{-}}{\sqrt{\tau }}\right] ^{\beta }\equiv 
\tilde{H}_{\alpha \beta }\otimes \left[ \frac{\tilde{\sigma}^{+}}{\sqrt{\tau 
}}\right] ^{\alpha }\left[ \frac{\tilde{\sigma}^{-}}{\sqrt{\tau }}\right]
^{\beta }
\end{equation*}
where 
\begin{eqnarray*}
\tilde{H}_{00} &=&H_{00}+\sqrt{\frac{p_{0}p_{1}}{\tau }}\theta ^{2}H_{10}+%
\sqrt{\frac{p_{0}p_{1}}{\tau }}\theta ^{\ast 2}H_{01}+\frac{p_{1}}{\tau }%
H_{11}; \\
\tilde{H}_{10} &=&p_{0}H_{10}-p_{1}H_{01}+\sqrt{\frac{p_{0}p_{1}}{\tau }}%
\theta ^{2}H_{11}; \\
\tilde{H}_{01} &=&p_{0}H_{01}-p_{1}H_{10}+\sqrt{\frac{p_{0}p_{1}}{\tau }}%
\theta ^{\ast 2}H_{11}; \\
\tilde{H}_{11} &=&\left( p_{0}-p_{1}\right) H_{11}-\sqrt{\tau p_{0}p_{1}}%
\left( \theta ^{2}+\theta ^{\ast 2}\right) H_{01}-\sqrt{\tau p_{0}p_{1}}%
\left( \theta ^{2}+\theta ^{\ast 2}\right) H_{10}.
\end{eqnarray*}

Let us now adopt $\tilde{e}_{0}$ as the new ground state and stabilizing
vector. We set $\tilde{\Phi}^{\tau }=\tilde{e}_{0}\otimes \tilde{e}%
_{0}\otimes \tilde{e}_{0}\otimes \cdots $. Clearly we would like to use the
tilded variables in place of the un-tilded ones however there is an
explosion problem associated with the coefficients $\tilde{H}_{\alpha \beta
} $. One way to resolve this is to allow the state to depend on $\tau $ by
taking 
\begin{equation}
p_{1}=\gamma ^{2}\tau +O\left( \tau ^{2}\right) .  \label{scale}
\end{equation}
In this case we have the asymptotic behaviour 
\begin{eqnarray*}
\tilde{H}_{00} &=&H_{00}+\gamma \theta ^{2}H_{10}+\gamma \theta ^{\ast
2}H_{01}+\gamma ^{2}H_{11}+O\left( \tau \right) ; \\
\tilde{H}_{10} &=&H_{10}+\gamma \theta ^{2}H_{11}+O\left( \tau \right) ; \\
\tilde{H}_{01} &=&H_{01}+\gamma \theta ^{\ast 2}H_{11}+O\left( \tau \right) ;
\\
\tilde{H}_{11} &=&H_{11}+O\left( \tau \right) .
\end{eqnarray*}
Here the continuous limit follows by using the above forms for the $\tilde{H}%
_{\alpha \beta }$ and ignoring the $O\left( \tau \right) $ terms. The
scaling used in (\ref{scale}) is necessary if we wish to obtain a Gaussian
limit for the collective operators: this is related to the notion of
macroscopic states in statistical mechanics \cite{Dubin}.

\end{document}